\documentclass[aps,prd,groupedaddress,showpacs]{revtex4}
\usepackage{graphicx}
\usepackage{dcolumn}
\usepackage{bm}
\begin{document}

\title{Lightest scalar and tensor resonances in \boldmath $\gamma\gamma\to\pi\pi$
after the Belle experiment}

\author{N.~N.~Achasov}
\email[]{achasov@math.nsc.ru}
\author{G.~N.~Shestakov}
\email[]{shestako@math.nsc.ru}

\altaffiliation{} \affiliation{Laboratory of  Theoretical Physics,
S.L. Sobolev Institute for Mathematics, 630090, Novosibirsk, Russia}

\date{\today}

\begin{abstract}
New high statistics Belle data on the $\gamma\gamma\to\pi^+\pi^-$
reaction cross section measured in the range of pion-pair invariant
masses $\sqrt{s}$ between 0.8 GeV and 1.5 GeV are analyzed to
clarify the current situation around the $\sigma(600)$, $f_0(980)$,
and $f_2(1270)$ resonances in $\gamma\gamma$ collisions. The present
analysis shows that the direct coupling constants of the
$\sigma(600)$ and $f_0(980)$ resonances to $\gamma\gamma$ are small,
and the $\sigma(600)\to\gamma\gamma$ and $f_0(980)\to\gamma\gamma$
decays are four-quark transitions caused by the $\pi^+\pi^-$ and
$K^+K^-$ loop mechanisms, respectively. The chiral shielding of the
$\sigma(600)$ resonance takes place in the reactions
$\gamma\gamma\to\pi\pi$ as well as in $\pi\pi$ scattering. Some
results of a simultaneous description of the
$\gamma\gamma\to\pi^+\pi^-$ and $\gamma\gamma\to\pi^0\pi^0$ Belle
data are also presented. In particular, the following tentative
estimate of the $f_2(1270)\to\gamma\gamma$ decay width is obtained:
$\Gamma_{f_2\to\gamma\gamma}(m^2_{f_2})\approx 3.68$ keV.
\end{abstract}

\maketitle

\section{INTRODUCTION}

Recently, the Belle Collaboration performed precise measurements of
the $\gamma\gamma\to\pi^+\pi^-$ reaction cross section for pion-pair
invariant masses $\sqrt{s}$ ranging from 0.8 to 1.5 GeV
\cite{B1,B2}. Owing to the huge statistics and good energy
resolution, a clear signal from the $f_0(980)$ resonance was first
discovered with the Belle detector. Evidences for the $f_0(980)$
production in $\gamma\gamma$ collisions obtained in a series of
previous measurements \cite{E1,E2,E3,E4,E5,E6} were essentially less
conclusive. The $f_0(980)$ signal observed in the Belle experiment
turned out to be rather small. This feature is in good qualitative
agreement with the prediction of the four-quark model \cite{ADS1}. A
detailed analysis of the preliminary Belle data \cite{B1} in the
$f_0(980)$ region was performed in Ref. \cite{AS1}. In particular,
it was found \cite{AS1} that the magnitude and shape of the
$f_0(980)$ peak observed in the $\gamma\gamma\to\pi^+\pi^-$ reaction
cross section excellently agree with the $K^+K^-$ loop mechanism of
the $f_0(980)$ production, $\gamma\gamma\to K^+K^-\to
f_0(980)\to\pi^+\pi^-$. This result is one of many in favor of the
four-quark nature of the $f_0(980)$ state \cite{AKS,A1}.

In this paper we clarify the current situation around the
$\sigma(600)$, $f_0(980)$, and $f_2(1270)$ resonances in the
reactions $\gamma\gamma\to\pi\pi$ by analyzing the final Belle data
\cite{B2} on the $\gamma\gamma\to\pi^+\pi^-$ cross section in the
region $0.8\leq\sqrt{s}\leq1.5$ GeV, together with the Crystal Ball
data \cite{E1,E5} on the reaction $\gamma\gamma\to\pi^0\pi^0$ for
$2m_\pi<\sqrt{s}<1.6$ GeV. As the first step, in Sec. II, the pure
Born cross sections and those involving the $S$ wave Born
contributions modified by strong final-state interactions are
compared with the available data on the reactions
$\gamma\gamma\to\pi\pi$ \cite{B2,E1,E2,E4,E5}. Such a comparison is
very useful because it enables us to gain some idea of the scope and
shape of other possible contributions to the cross sections. The
scheme taking into account the $S$ wave final-state interactions
between pions and kaons, in which the $\sigma(600)$ and $f_0(980)$
resonances take part, is presented in Sec. III. In this scheme we
essentially use the results of the simultaneous analysis of the data
on the $\phi\to\pi^0\pi^0\gamma$ decay, $\pi\pi$ scattering, and
reaction $\pi\pi\to K\bar K$ \cite{AK1}, as well as the results of
the previous analyses of the $\gamma\gamma\to\pi\pi$ reaction
mechanisms \cite{AS1,AS2}. In Secs. IV and V, tentative values of
the direct coupling constants of the $\sigma(600)$ and $f_0(980)$
resonances to $\gamma\gamma$ are determined from the Belle and
Crystal Ball data. It is important that these constants turn out to
be small, and consequently, the $\sigma(600)\to\gamma\gamma$ and
$f_0(980)\to\gamma\gamma$ decays are, in fact, the four-quark
transitions caused by the $\pi^+\pi^-$ and $K^+K^-$ loop mechanisms,
respectively. The chiral shielding of the $\sigma(600)$ resonance
takes place in the reactions $\gamma\gamma\to\pi\pi$ \cite{AS2} as
well as in $\pi\pi$ scattering \cite{AKS,A1,AK1,AS3}. In addition,
in Sec. V, we make some comments on the difficulties of interpreting
the experimental measurements for $\gamma\gamma\to\pi^0\pi^0$
production in the $f_2(1270)$ resonance region. Here we also point
to a possible way out and announce some preliminary results of a
simultaneous description of the $\gamma\gamma\to\pi^+\pi^-$ data and
the latest, very high statistics Belle data on the
$\gamma\gamma\to\pi^0\pi^0$ reaction cross section. The estimates
for the $f_2(1270)\to\gamma\gamma$ decay width are also given.
Conclusions are shortly formulated in Sec. VI.

\section{DATA ON {\boldmath
$\gamma\gamma\to\pi\pi$} AND THE BORN CONTRIBUTIONS}

The Mark II \cite{E2}, CELLO \cite{E4}, and Belle \cite{B2,B3} data
on the cross section for the reaction $\gamma\gamma\to\pi^+\pi^-$
are shown in Fig. 1(a). All the data correspond to a limited angular
range of the registration of the charged pions, such that $|\cos
\theta|\leq0.6$, where $\theta$ is the polar angle of the produced
$\pi^\pm$ meson in the center-of-mass system of two initial photons.
The Belle data are represented with statistical errors only, the
relative values of which are approximately equal to
0.5\,\%--1.5\,\%. The $\sqrt{s}$ bin size in the Belle experiment
has been chosen to be 5 MeV, with the mass resolution of about 2
MeV.

Figure 1(a) also represents the comparison between the data and
theoretical curves corresponding to the pure Born cross sections for
the process $\gamma\gamma\to\pi^+\pi^-$ for $|\cos \theta|\leq0.6$:
the total integrated cross section
$\sigma^{\mbox{\scriptsize{Born}}}$ and the integrated cross
sections $\sigma^{\mbox{\scriptsize{Born}}}_\lambda$, where
$\lambda=0$ or 2 is the absolute value of the photon helicity
difference. According the Low theorem, the Born contributions give
the exact physical amplitude of the crossing reaction
$\gamma\pi^\pm\to\gamma\pi^\pm$ near its threshold. If strong
interactions of pions near the $\pi\pi$ threshold are not too large
(this is the case owing to chiral symmetry), then the Born
contributions have to dominate near the threshold of the reaction
$\gamma\gamma\to\pi^+\pi^-$ as well. As is seen from Fig. 1(a), this
expectation does not contradict the data existing in the threshold
region, though their errors are very large as yet. Moreover, the
Born amplitudes can be used as a quite reasonable approximation of
the primary background (nonresonance) contributions in the reaction
$\gamma\gamma\to\pi^+\pi^-$ up to the $f_2(1270)$ resonance region,
as well as a reasonable foundation for constructing the amplitudes
involving the strong final-state interactions; see, for example,
\cite{AS1,AS2,Me,Ly,Jo,MP,Pe}. The pure Born contributions have the
following features. First, $\sigma^{\mbox{\scriptsize{Born}}}$
reaches its maximum at $\sqrt{s}\approx0.3$ GeV, in the vicinity of
which $\sigma^{\mbox{\scriptsize{Born}}}\approx\sigma^{
\mbox{\scriptsize{Born} }}_0$. However,
$\sigma^{\mbox{\scriptsize{Born}}}_0$ involving the $S$ wave
contribution decreases rapidly with increasing $\sqrt{s}$ so that
$\sigma^{\mbox{\scriptsize{Born}}}$ is fully dominated by the
contribution from $\sigma^{\mbox{\scriptsize{Born}}}_2$ for
$\sqrt{s}>0.5$ GeV; see Fig. 1(a). Second, though
$\sigma^{\mbox{\scriptsize{Born}}}_2$ is determined essentially by
the $D$ wave contribution with $\lambda=2$, the differential cross
section $d\sigma^{\mbox{\scriptsize{Born}}}/d|\cos\theta|$ in the
region $|\cos \theta|\leq0.6$ changes very weakly (see Fig. 1(b)),
and it becomes more and more flat in this region of $|\cos \theta|$
as $\sqrt{s}$ increases. Therefore, in analyzing the data
corresponding to this angular range, one should be very careful to
make a definite conclusion about the partial wave structure of the
smooth background. In other words, the above example hints that the
assumption of the $S$ wave dominance \cite{B3} may be completely
unjustified in reality.

Figure 1(c) shows only the Belle data in the region $\sqrt{s}>0.85$
to illustrate clearly the discovered signal from the $f_0(980)$
resonance. The visible height of the $f_0(980)$ peak is of about 15
nb over the overall smooth background of the order of 100 nb, and
its visible (effective) width is of about $30-35$ MeV. It is natural
to consider that the large background under the $f_0(980)$ resonance
is mainly caused by the contributions from the Born amplitude with
$\lambda=2$ and the $f_2(1270)$ resonance production amplitude also
with $\lambda=2$. Figure 1(c) shows the theoretical curves for the
total integrated cross section
$\sigma=\sigma_0+\sigma^{\mbox{\scriptsize{Born}}}_2$ and the
integrated ones $\sigma_0$ and
$\sigma^{\mbox{\scriptsize{Born}}}_2$. Here $\sigma_0$ is the
$\gamma\gamma\to\pi^+\pi^-$ cross section with $\lambda=0$ in which
the contributions of the $S$ wave Born amplitude are modified by the
strong final-state interactions. All the higher partial waves, $D$,
$G$, etc., in $\sigma_0$ are taken in the Born approximation
\cite{AS1}. The above modification leads to a signal from the
$f_0(980)$ resonance in $\sigma_0$ whose magnitude and shape are in
surprising agreement with the Belle data; see Fig. 1(c). An explicit
form for $\sigma_0$ will be given below. The comparison of the
curves in Figs. 1(c) and 1(a) shows that the $S$ wave contribution
to $\sigma(\gamma\gamma\to\pi^+\pi^-; |\cos\theta|\leq0.6)$ is small
for $\sqrt{s}>0.5$ in any case. Certainly, the $f_2(1270)$ resonance
contribution is an important component needed for the description of
the Belle data in the whole region of $\sqrt{s}$ from 0.8 to 1.5
GeV.

\begin{figure} \includegraphics{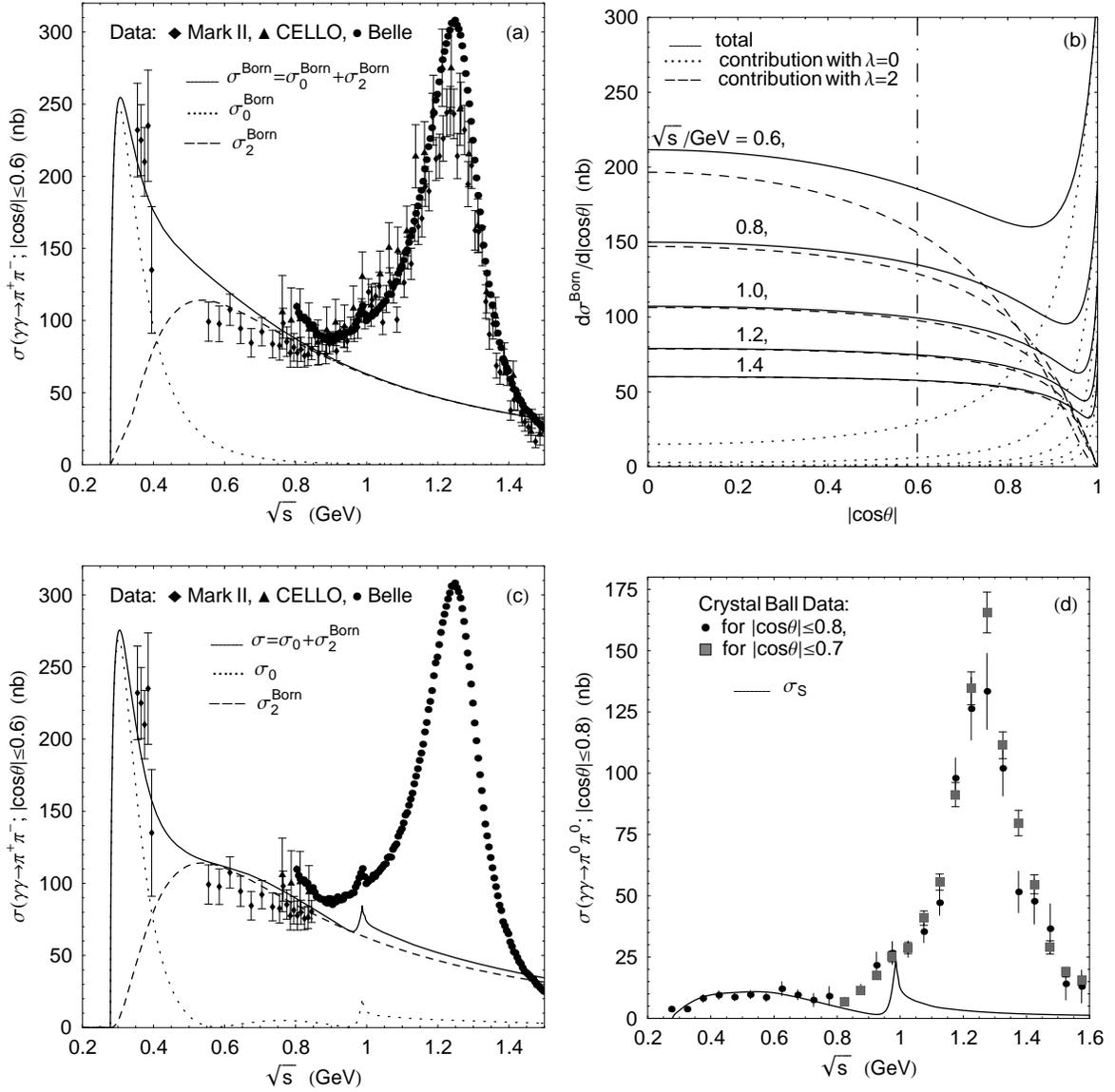} \caption {The data from Mark II
\cite{E1}, CELLO \cite{E4}, and Belle \cite{B2,B3} on the cross
section for $\gamma\gamma\to\pi^+\pi^-$, and from Crystal Ball
\cite{E2,E5} on the cross section for $\gamma\gamma\to\pi^0\pi^0$.
The theoretical curves correspond to the pure Born cross sections,
as well as the Born cross sections modified for strong final-state
interactions in the $S$ wave. See the text for details.}
\end{figure}

The incorporation of final-state interactions in the $S$ wave Born
amplitude $\gamma\gamma\to\pi^+\pi^-$ leads to a striking prediction
for the $S$ wave amplitude $\gamma\gamma\to\pi^0\pi^0$ \cite{AS1}.
Figure 1(d) demonstrates the comparison of the $S$ wave partial
cross section for the reaction $\gamma\gamma\to\pi^0 \pi^0$,
$\sigma_S$, calculated in the way outlined above, with the Crystal
Ball data \cite{E2,E5}. Since $\sigma_S$ does not contain any
fitting parameter, the agreement with experiment in the region
$2m_\pi\leq\sqrt{s}\leq0.8$ GeV is quite reasonable. Note that, in
the Crystal Ball experiments \cite{E2,E5}, the
$\gamma\gamma\to\pi^0\pi^0$ cross section was scanned with a
50-MeV-wide step, and therefore the narrow $f_0(980)$ structure
could not be resolved. It is also clear that in the reaction
$\gamma\gamma\to\pi^0\pi^0$, as well as in
$\gamma\gamma\to\pi^+\pi^-$, the $f_2(1270)$ resonance manifestation
domain begins for $\sqrt{s}>0.8$ GeV. The latest, high statistics
Belle data on the $\gamma\gamma\to\pi^0\pi^0$ reaction cross section
in the region $0.73<\sqrt{s}<1.5$ GeV are presented and discussed
below in Sec. V.

\section{{\boldmath $S$} WAVE FINAL-STATE INTERACTIONS}

In the field theory approach, one has the following $S$ wave
$\gamma\gamma\to\pi\pi$ amplitudes, satisfying the unitarity
condition and involving the electromagnetic Born contributions
``dressed'' (modified) by strong final-state interactions
\cite{AS1,AS2,Me,AG1},
\begin{eqnarray}\label{MSpm}& M_S(\gamma\gamma\to\pi^+\pi^-;s)
=M^{\mbox{\scriptsize{Born}}}_S(s)+8\alpha
I_{\pi^+\pi^-}(s)\,T_{\pi^+\pi^-\to\pi^+\pi^-}(s) & \nonumber \\ &
+8\alpha I_{K^+K^-}(s)\,T_{K^+K^- \to\pi^+\pi^-}(s) & \nonumber \\ &
=(\mbox{for\ } 2m_\pi\leq\sqrt{s}\leq2m_K)=
\frac{2}{3}e^{i\delta^0_0(s)}\left
\{M^{\mbox{\scriptsize{Born}}}_S(s)\cos\delta^0_0(s)
+\frac{8\alpha}{\rho_{\pi^+}(s)}\,\mbox{Re}[I_{\pi^+\pi^-}(s)]
\sin\delta^0_0(s)\right. & \nonumber\\ & \left.\pm12\alpha
I_{K^+K^-}(s)|T_{K^+K^- \to\pi^+\pi^-}(s)|\right\} & \nonumber \\
& + \frac{1}{3}e^{i\delta^2_0(s)}\left
\{M^{\mbox{\scriptsize{Born}}}_S(s)\cos\delta^2_0(s)+
\frac{8\alpha}{\rho_{\pi^+}(s)}\,
\mbox{Re}[I_{\pi^+\pi^-}(s)]\sin\delta^2_0(s)\right\}\,, &
\end{eqnarray}
\begin{eqnarray}\label{MS00}& M_S(\gamma\gamma\to\pi^0\pi^0;s)
=8\alpha I_{\pi^+\pi^-}(s)\,T_{\pi^+\pi^-\to\pi^0\pi^0}(s) +8\alpha
I_{K^+K^-}(s)\,T_{K^+K^- \to\pi^0\pi^0}(s) & \nonumber \\ &
=(\mbox{for\ } 2m_\pi\leq\sqrt{s}\leq2m_K)=
\frac{2}{3}e^{i\delta^0_0(s)}\left
\{M^{\mbox{\scriptsize{Born}}}_S(s)\cos\delta^0_0(s)
+\frac{8\alpha}{\rho_{\pi^+}(s)}\,\mbox{Re}[I_{\pi^+\pi^-}(s)]
\sin\delta^0_0(s)\right. & \nonumber\\ & \left.\pm12\alpha
I_{K^+K^-}(s)|T_{K^+K^- \to\pi^0\pi^0}(s)|\right\} & \nonumber \\
& - \frac{2}{3}e^{i\delta^2_0(s)}\left
\{M^{\mbox{\scriptsize{Born}}}_S(s)\cos\delta^2_0(s)+
\frac{8\alpha}{\rho_{\pi^+}(s)}\,
\mbox{Re}[I_{\pi^+\pi^-}(s)]\sin\delta^2_0(s)\right\}\,, &
\end{eqnarray}
where $M^{\mbox{\scriptsize{Born}}}_S(s)$ is the $S$ wave Born
amplitude of the process $\gamma\gamma$\,$\to$\,$\pi^+\pi^-$; for
$s\geq4m_\pi^2$,
\begin{equation}\label{MSB}M^{\mbox{\scriptsize{Born}}}_S(s)=\frac{16\pi\alpha m^2
_{\pi^+}}{s\rho_{\pi^+}(s)}\,\ln\frac{1+\rho_{\pi^+}(s)}{1-\rho_{\pi^+}(s)}=\frac
{8\alpha}{\rho_{\pi^+}(s)}\mbox{Im}I_{\pi^+\pi^-}(s)\,,
\end{equation}
$\rho_{\pi^+}(s)=(1-4m^2_{\pi^+}/s)^{1/2}$, $\,\delta^0_0(s)$ and
$\delta^2_0(s)$ are the phase shifts of the $S$ wave $\pi\pi$
scattering amplitudes with isospin $I=0$ and 2, respectively (see
below for details), and $\alpha=1/137$. The second equalities in
Eqs. (\ref{MSpm}) and (\ref{MS00}) are valid just in the elastic
region, practically for $2m_\pi\leq\sqrt{s}\leq2m_K$, and clearly
demonstrate the fulfillment of the Watson theorem for the $S$ wave
$\gamma\gamma\to\pi\pi$ amplitudes with definite isospin. The
function $I_{K^+K^-}(s)$ is given by
\begin{eqnarray}\label{IKK} I_{K^+K^-}(s)=\left\{\begin{array}{ll}
\frac{m^2_{K^+}}{s}\left[\pi+i\ln\frac{1+\rho_{K^+}(s)}
{1-\rho_{K^+}(s)}\right]^2-1\,,\ \ & s\geq4m^2_{K^+}\,, \\
\frac{m^2_{K^+}}{s}[\pi-2\arctan|\rho_{K^+}(s)|]^2-1\,,\ \ & 0\leq
s\leq4m^2_{K^+}\,, \end{array}\right.
\end{eqnarray}
$\rho_{K^+}(s)=(1-4m^2_{K^+}/s)^{1/2}$ for $s\geq4m^2_{K^+}$ and
$\rho_{K^+}(s)\to i|\rho_{K^+}(s)|$ if $0\leq s\leq4m^2_{K^+}$.
$I_{\pi^+\pi^-}(s)$ results from $I_{K^+K^-}(s)$ by replacing
$m_{K^+}$ and $\rho_{K^+}(s)$ by $m_{\pi^+}$ and $\rho_{\pi^+}(s)$,
respectively. Finally, $T_{\pi^+\pi^-\to\pi^+\pi^-}(s)$,
$T_{\pi^+\pi^-\to\pi^0\pi^0}(s)$, and
$T_{K^+K^-\to\pi^+\pi^-}(s)$\,=\,$T_{K^+K^-\to\pi^0\pi^0}(s)$ are
the $S$ wave amplitudes of hadronic reactions indicated in their
subscripts. Equations (\ref{MSpm}) and (\ref{MS00}) assume that the
amplitudes $T_{\pi^+\pi^-\to\pi\pi}(s)$ and $T_{K^+K^-\to\pi\pi}(s)$
lie on the mass shell in the rescattering loops
$\gamma\gamma$\,$\to$\,$\pi^+\pi^-$\,$\to$\,$\pi\pi$ and
$\gamma\gamma$\,$\to$\,$K^+K^-$\,$\to$\,$\pi\pi$. The functions
$I_{\pi^+\pi^-}(s)$ and $I_{K^+K^-}(s)$ are the attributes of the
triangle diagrams
$\gamma\gamma$\,$\to$\,$\pi^+\pi^-$\,$\to$\,$\sigma$,\,$f_0$ and
$\gamma\gamma$\,$\to$\,$K^+K^-$\,$\to$\,$\sigma$,\,$f_0$ (or any
other scalars). Thus, the amplitudes in Eqs. (\ref{MSpm}) and
(\ref{MS00}) represent the $S$ wave
$\gamma\gamma$\,$\to$\,$\pi^+\pi^-$ and
$\gamma\gamma$\,$\to$\,$K^+K^-$ Born contributions modified by the
strong elastic and inelastic final-state interactions.

The helicity-0 cross section integrated over the range
$|\cos\theta|\leq Z_0<1$ and involving the amplitude
$M_S(\gamma\gamma\to\pi^+\pi^-;s)$ and the pure Born higher partial
wave amplitudes can be written in the form
$$\sigma_{\lambda=0}(\gamma\gamma\to\pi^+\pi^-,|\cos\theta|\leq Z_0)=
\frac{\rho_{\pi^+}(s)}{32\pi
s}\Biggl\{Z_0|\widetilde{A}_S(s)|^2+C\,\mbox{Re}[\widetilde{A}_S(s)]
\Biggr. $$ \begin{equation}\label{CS0pm}\left.\times
\frac{1}{\rho_{\pi^+}(s)}\,\ln\frac{1+Z_0\rho_{\pi^+}(s)}{1-Z_0\rho_{\pi^+}(s)}+
C^2\left[\frac{Z_0/2}{1-Z^2_0\rho^2_{\pi^+}(s)}+\frac{1}{4\rho_{\pi^+}(s)}
\,\ln\frac{1+Z_0\rho_{\pi^+}(s)}{1-Z_0\rho_{\pi^+}(s)}\right]\right\}\,,
\end{equation} where $\widetilde{A}_S(s)=M_S(
\gamma\gamma\to\pi^+\pi^-;s)-M^{\mbox{\scriptsize{Born}}}_S(s)$ [see
Eq. (\ref{MSpm})] and $C=32\pi\alpha m^2_{\pi^+}/s$. The cross
section $\sigma_0$, shown, for example, in Fig. 1(c) by the dotted
curve (see also the solid curve in this figure and the next section
for details), is given by
\begin{equation}\label{CS0}\sigma_0=\sigma_{\lambda=0}(\gamma\gamma\to\pi^+\pi^-,
|\cos\theta|\leq 0.6)\,.\end{equation} Similarly, the $S$ wave cross
section for the reaction $\gamma\gamma\to\pi^0\pi^0$ is given by
\begin{equation}\label{CSS00}\sigma_S(\gamma\gamma\to\pi^0\pi^0)=
\frac{\rho_{\pi^+}(s)}{64\pi s}\left
|M_S(\gamma\gamma\to\pi^0\pi^0;s)\right |^2\,,
\end{equation}
see Eq. (\ref{MS00}). The cross section $\sigma_S$ shown in Fig.
1(d) corresponds to $0.8\,\sigma_S(\gamma\gamma\to\pi^0\pi^0)$.

To construct $\sigma_0$ and $\sigma_S$, the amplitudes
$T_{\pi^+\pi^-\to\pi^+\pi^-}(s)$, $T_{\pi^+\pi^-\to \pi^0\pi^0}(s)$,
and $T_{K^+K^-\to\pi^+\pi^-}(s)=T_{\pi^+\pi^-\to K^+K^-}(s)$ need to
be known. They are related to the $S$ wave $\pi\pi$ scattering
amplitude $T^I_0(s)$, the phase shift $\delta^I_0(s)$, and
inelasticity $\eta^I_0(s)$ with definite isospin $I=0,2$ in the
conventional way:
\begin{equation}\label{Tpipi}  T_{
\pi^+\pi^-\to\pi^+\pi^-}(s)=[2T^0_0(s)+T^2_0(s)]/3,\qquad
T_{\pi^+\pi^-\to\pi^0\pi^0}(s)=2[T^0_0(s)-T^2_0(s)]/3\,,\end{equation}
\begin{equation}\label{TI0}
T^I_0(s)=\{\eta^I_0(s)\exp[2i\delta^I_0(s)]-1\}/[2i\rho_{\pi^+}(s)]\,,\end{equation}
\begin{equation}\label{eta00} \eta^0_0(s)=\sqrt{1-4\rho_{K^+}\rho_{\pi^+}
(3/2)|T_{\pi^+\pi^-\to K^+K^- }(s)|^2\theta(s-4m^2_{K^+})-...}\ .
\end{equation}
Dots in $\eta^0_0(s)$ denote the contributions from other inelastic
channels $\pi^+\pi^-\to K^0\bar K^0$, $\pi^+\pi^-\to\eta\eta$, etc.
For $4m_\pi^2\leq s\leq4m^2_K$, the amplitude $T_{\pi^+\pi^-\to
K^+K^- }(s)=\pm e^{i\delta^0_0(s)}|T_{\pi^+\pi^-\to K^+K^- }(s)|$
for the $4\pi$ channel contribution is small \cite{AS1,AK1}. Note
that the plus sign is realized here according to Ref. \cite{AK1}. We
also set $\eta^2_0(s)=1$ for all $s$ of interest.

A parametrization of the amplitudes $T^0_0(s)$ and
$T_{K^+K^-\to\pi^+\pi^-}(s)$ has been thoroughly described in Ref.
\cite{AK1}. It has been used for the simultaneous analysis of the
data on the $\pi^0\pi^0$ mass distribution in the
$\phi\to\pi^0\pi^0\gamma$ decay, $\pi\pi$ scattering in the region
$2m_\pi\leq\sqrt{s}<1.6$ GeV, and the reaction $\pi\pi\to K\bar K$
\cite{AK1}. The key idea of this parametrization is that the
amplitude $T^0_0(s)$ incorporates the contributions from the mixed
$\sigma(600)$ and $f_0(980)$ resonances and from the nonresonant
background having a large negative phase which hides the
$\sigma(600)$ meson \cite{AKS,A1,AK1}. Originally, the presence of
such a background in $\pi\pi$ scattering was established in the
linear $\sigma$ model \cite{AS3}. It has been made clear that
shielding of wide, lightest scalar mesons in chiral dynamics is very
natural. As for $\gamma\gamma$ interactions, Eqs. (1) and (2)
transfer the chiral shielding effect of the $\sigma(600)$ from
$\pi\pi$ scattering to the $\gamma\gamma\to\pi\pi$ reaction
amplitudes \cite{AS2}. The shielding of the $\sigma$ meson takes
place in the $\gamma\gamma\to\pi\pi$ amplitudes for the strong
destructive interference between the resonance and background
contributions as in the $\pi\pi\to\pi\pi$ amplitudes. This was first
demonstrated in the frame of the $SU_L(2)\times SU_R(2)$ linear
$\sigma$ model in Ref. \cite{AS2}. If such a shielding was absent,
then the cross section $\sigma_S$ would reach approximately 100 nb
just above the $\pi^0\pi^0$ threshold, owing to the $\pi^+\pi^-$
loop mechanism of the $\sigma(600)\to\gamma\gamma$ decay \cite{AS2},
instead of about 10 nb as in experiment  [see Fig. 1(d)].

We now return to the parametrization of the strong amplitudes. In
terms of the mixed $\sigma(600)$ and $f_0(980)$ resonances and the
necessary background contributions, the explicit form of the
amplitudes $T^0_0(s)$ and $T_{K^+K^-\to\pi^+\pi^-}(s)$ is given by
\cite{AK1}
\begin{equation}\label{T00}
T^0_0(s)=\frac{\eta^0_0(s)e^{2i\delta^0_0
(s)}-1}{2i\rho_{\pi^+}(s)}=T^{\pi\pi}_B(s)+
e^{2i\delta^{\pi\pi}_{B}(s)}
T^{\pi\pi}_{\mbox{\scriptsize{res}}}(s)\,,\end{equation}
\begin{equation}\label{TKpi}T_{K^+K^-\to\pi^+\pi^-}(s)=
e^{i[\delta^{\pi\pi}_{B}(s)+\delta^{K\bar K}_{B}(s)]}
T^{K^+K^-\to\pi^+\pi^-}_{\mbox{\scriptsize{res}}}(s)\,,
\end{equation} where $\delta^{\pi\pi}_{B}(s)$ and
$\delta^{K\bar K}_{B}(s)$ are the phases of the elastic background
in the $I=0$ $S$ wave $\pi\pi$ and $K\bar K$ channels, respectively,
$T^{\pi\pi}_B(s)=\{\exp[2i\delta^{\pi\pi}_{B}(s)]
-1\}/[2i\rho_{\pi^+}(s)]$ is the $I=0$ $S$ wave $\pi\pi$ background
amplitude, the amplitudes of the $\sigma(600)-f_0(980)$ resonance
complex are
$$T^{\pi\pi}_{\mbox{\scriptsize{res}}}(s)=\frac{\eta^0_0(s)e^{2i\delta
_{\mbox{\scriptsize{res}}}(s)}-1}{2i\rho_{\pi^+}(s)}$$
\begin{equation}\label{Tpipir}=\frac{3}{32\pi}\frac{g_{\sigma\pi^+\pi^-}
[D_{f_0}(s)g_{\sigma\pi^+\pi^-}+\Pi_{f_0\sigma}(s)g_{f_0\pi^+
\pi^-}]+g_{f_0\pi^+\pi^-}[D_{\sigma}(s)g_{f_0\pi^+\pi^-}+
\Pi_{f_0\sigma}(s)g_{\sigma\pi^+
\pi^-}]}{D_\sigma(s)D_{f_0}(s)-\Pi^2_{f_0\sigma}(s)}\,,
\end{equation} and
$$T^{K^+K^-\to\pi^+\pi^-}_{\mbox{\scriptsize{res}}}(s)$$
\begin{equation}\label{TKpir}=\frac{1}{16\pi}\frac{g_{\sigma K^+K^-}
[D_{f_0}(s)g_{\sigma\pi^+\pi^-}+\Pi_{f_0\sigma}(s)g_{f_0\pi^+
\pi^-}]+g_{f_0K^+K^-}[D_{\sigma}(s)g_{f_0\pi^+\pi^-}+
\Pi_{f_0\sigma}(s)g_{\sigma\pi^+
\pi^-}]}{D_\sigma(s)D_{f_0}(s)-\Pi^2_{f_0\sigma}(s)}\,,
\end{equation} and the phase shift
$\delta^0_0(s)=\delta^{\pi\pi}_B(s)+\delta_{\mbox{\scriptsize{res}}}(s)$.
We use for $\delta^{\pi\pi}_B(s)$, for propagators of the
$\sigma(600)$ and $f_0(980)$ resonances $1/D_\sigma(s)$ and
$1/D_{f_0}(s)$, and for the nondiagonal matrix element of the
polarization operator $\Pi_{f_0\sigma}(s)$, the expressions
presented in Ref. \cite{AK1} (see also Ref. \cite{FN1}). In the
accepted normalization the relation between the coupling constant
$g_{\sigma\pi^+\pi^-}$ and the corresponding partial decay width of
the $\sigma(600)$ is given by
$\Gamma_{\sigma\to\pi\pi}(s)=[3g_{\sigma\pi^+\pi^-}^2/
(32\pi)]\,\rho_{\pi^+}(s)$. Similar relations take place for the
$\sigma(600)$ decays into $K\bar K$, $\eta\eta$, $\eta\eta'$, and
$\eta'\eta'$, and for the $f_0(980)$ decays into $\pi\pi$, $K\bar
K$, $\eta\eta$, $\eta\eta'$, and $\eta'\eta'$. Remember that the
$\sigma(600)$ couples mainly to $\pi\pi$, $\eta\eta$, $\eta\eta'$,
and $\eta'\eta'$, and the $f_0(980)$ to $K\bar K$, $\eta\eta$,
$\eta\eta'$, and $\eta'\eta'$ \cite{AK1}.

The various fits corresponding to the different values of the
parameters in the strong amplitudes have been considered in Ref.
\cite{AK1}. All of these fits give excellent descriptions of a large
set of the data on the $\phi\to\pi^0\pi^0\gamma$ decay,
$\delta^0_0(s)$, and $\eta^0_0(s)$. The curves for $\sigma_0$ and
$\sigma$ in Fig. 1(c), and for $\sigma_S$ in Fig. 1(d), calculated
with the use of Eqs. (\ref{MSpm}), (\ref{MS00}), (\ref{CS0pm}) and
(\ref{CSS00}), correspond to fit 1 from Table I presented in Ref.
\cite{AK1}; see also Ref. \cite{FN2}. For the phase shift
$\delta^2_0(s)$, we take the parametrization of Ref. \cite{AS4}.

According Eqs. (\ref{MSpm}) and (\ref{MS00}), the
$\sigma(600)$\,$\to$\,$\gamma\gamma$ and
$f_0(980)$\,$\to$\,$\gamma\gamma$ decays are described by the
$\pi^+\pi^-$ and $K^+K^-$ loop diagrams,
Resonances\,$\to$\,($\pi^+\pi^-$,\,$K^+K^-$)\,$\to$\,$\gamma \gamma$
[$I_{\pi^+\pi^-}(s)$, $I_{K^+K^-}(s)$]. Consequently, they are the
four-quark transitions \cite{AS2}. Let us emphasize that there are
no free parameters specific for the reactions
$\gamma\gamma\to\pi\pi$ in Eqs. (\ref{MSpm}) and (\ref{MS00}), and
that the existing data are not indicative of the necessity of
introducing such parameters. Nevertheless, in the next section we
shall supplement Eqs. (\ref{MSpm}) and (\ref{MS00}) with the terms
involving unknown direct coupling constants of the $\sigma(600)$ and
$f_0(980)$ resonances to $\gamma\gamma$, and we shall attempt to
extract the values of these undoubtedly important physical
characteristics from the data. Note that some evidence for smallness
of these constants was obtained previously in Refs. \cite{AS1,AS2}
within the more simple models for the amplitudes
$\gamma\gamma\to\pi\pi$. Later, a similar conclusion about the
direct coupling of the $\sigma(600)$ to $\gamma\gamma$ was also
obtained in Ref. \cite{MM}.

\section{DIRECT COUPLINGS OF THE \boldmath $\sigma(600)$
AND \lowercase{$f_0(980)$} TO $\gamma\gamma$}

We now add to the right-hand side of Eq. (\ref{MSpm}) the amplitude
$M^{\mbox{\scriptsize{direct}}}_{\mbox{\scriptsize{res}}}
(\gamma\gamma\to\pi^+\pi^-;s)$ caused by the contribution from the
mixed $\sigma(600)$ and $f_0(980)$ resonances \cite{AK1} with the
direct coupling constants of the $\sigma(600)$ and $f_0(980)$ to
photons, $g^{(0)}_{\sigma\gamma\gamma}$ and
$g^{(0)}_{f_0\gamma\gamma}$,
$$M^{\mbox{\scriptsize{direct}}}_{\mbox{\scriptsize{res}}}
(\gamma\gamma\to\pi^+\pi^-;s)=s\,e^{i\delta^{\pi\pi}_B(s)}$$
\begin{equation}\label{Mrd}\times\frac{g^{(0)}_{\sigma\gamma
\gamma}[D_{f_0}(s)g_{\sigma\pi^+\pi^-}+\Pi_{f_0\sigma}(s)g_{f_0\pi^+
\pi^-}]+g^{(0)}_{f_0\gamma
\gamma}[D_{\sigma}(s)g_{f_0\pi^+\pi^-}+\Pi_{f_0\sigma}(s)g_{\sigma\pi^+
\pi^-}]}{D_\sigma(s)D_{f_0}(s)-\Pi^2_{f_0\sigma}(s)}\,.
\end{equation}
To the right-hand side of Eq. (\ref{MS00}), we also add the
amplitude
$M^{\mbox{\scriptsize{direct}}}_{\mbox{\scriptsize{res}}}(\gamma\gamma\to\pi^0\pi^0;s)$\,=\,$
M^{\mbox{\scriptsize{direct}}}_{\mbox{\scriptsize{res}}}(\gamma\gamma\to\pi^+\pi^-;s)$.
The factor $s$ in Eq. (\ref{Mrd}) is due to gauge invariance. The
above amplitude also satisfies the unitarity condition. For
$\sqrt{s}<2m_K$, its phase coincides with the $I=0$ $S$ wave
$\pi\pi$ phase shift
$\delta^0_0(s)=\delta^{\pi\pi}_B(s)+\delta_{\mbox{\scriptsize{res}}}(s)$.

Of the existing data only the Belle ones on the reaction
$\gamma\gamma\to\pi^+\pi^-$ in the vicinity of the $f_0(980)$ [Fig.
1(c)] and the Crystal Ball data on the reaction
$\gamma\gamma\to\pi^0\pi^0$ for $2m_\pi<\sqrt{s}<0.8$ GeV [Fig.
1(d)] may be sensitive to the coupling constants
$g^{(0)}_{\sigma\gamma\gamma}$ and $g^{(0)}_{f_0\gamma\gamma}$.
Therefore, to estimate $g^{(0)}_{\sigma\gamma\gamma}$ and
$g^{(0)}_{f_0\gamma\gamma}$ we perform a simultaneous fit to the
Crystal Ball data in the above region of $\sqrt{s}$ and the Belle
data for $0.85<\sqrt{s}<1.15$ GeV. Inclusion of the
$\gamma\gamma\to\pi^+\pi^-$ data from a sufficiently wide region
around the narrow $f_0(980)$ resonance in the fit is dictated by the
following circumstance. The $f_0(980)$ peak is observed in the total
cross section
$\sigma(\gamma\gamma\to\pi^+\pi^-;|\cos\theta|\leq0.6)=\sigma_0+\sigma_2$
under the very large, noncoherent, smooth background caused by the
contribution of the cross section with $\lambda=2$, i.e.,
$\sigma_2=\sigma_{\lambda=2}(\gamma\gamma\to\pi^+\pi^-,|\cos\theta|\leq
0.6)$. It is natural that the $\sqrt{s}$ dependence of this
background in the $f_0(980)$ region can be fixed more or less
reliably only with the use of the data outside this region.
Certainly, $\sigma_2$ is dominated by the Born and  $f_2(1270)$
resonance contributions. However, first we use a purely
phenomenological approximation of $\sigma_2$ with a 4th order
polynomial in $\sqrt{s}$ in the region $0.85<\sqrt{s}<1.15$ GeV. In
addition, to obtain a correct fit to the Belle data having the
finest step in $\sqrt{s}$, we allow the $f_0(980)$ resonance mass,
$m_{f_0}$, to be a free parameter. We fix the values of the other
parameters in the strong amplitudes in accordance with fit 1 from
Ref. \cite{AK1}.

\begin{figure} \includegraphics{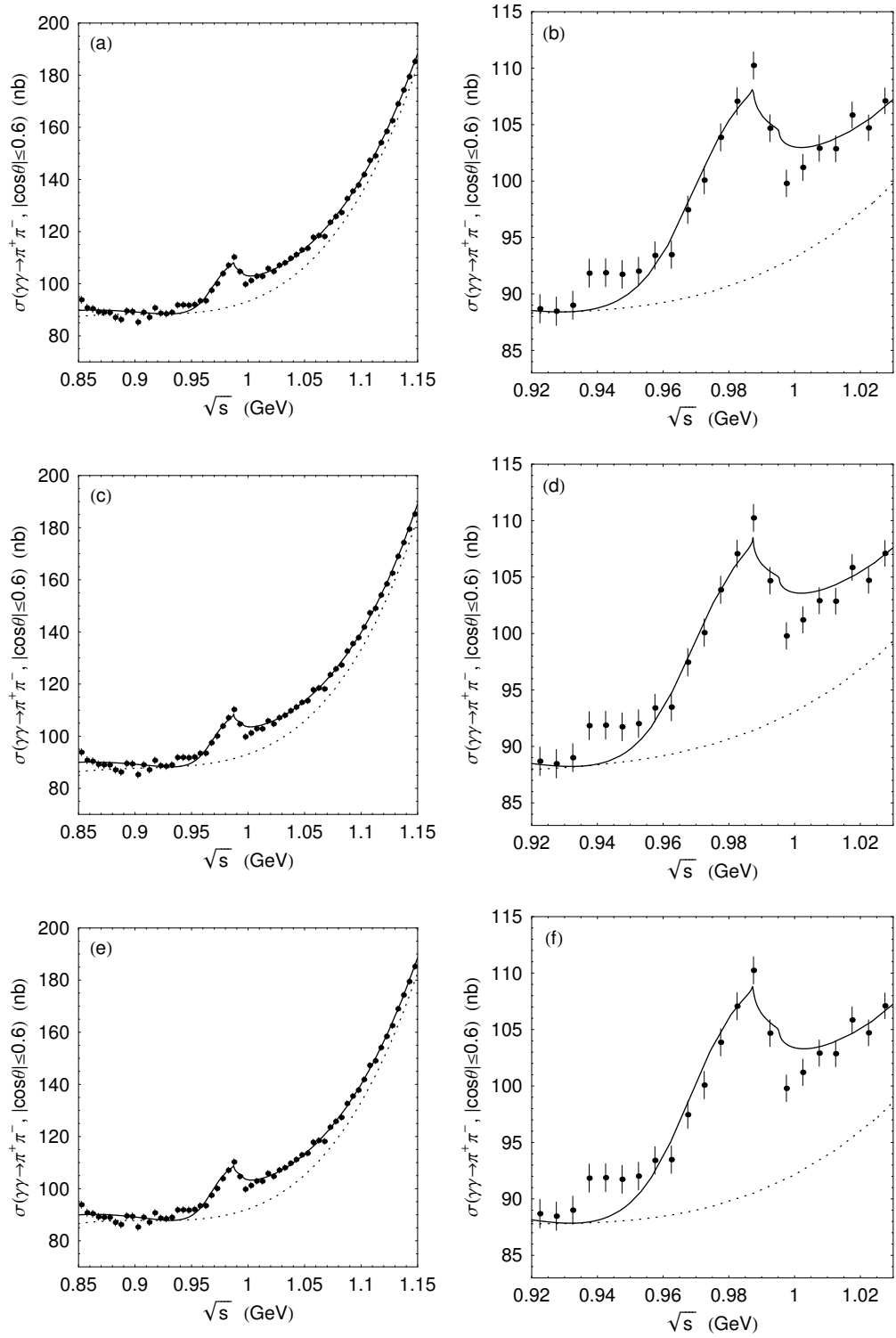} \caption {The results of
the three fits pertaining to the Belle data in the vicinity of the
$f_0(980)$. They show that the direct couplings of the $\sigma(600)$
and $f_0(980)$ resonances to $\gamma\gamma$ are small. The solid and
dotted curves correspond to the cross sections
$\sigma(\gamma\gamma\to\pi^+\pi^-;|\cos\theta|\leq0.6)=\sigma_0+\sigma_2$
and $\sigma_2$, respectively. The right-hand plots emphasize the
region of the $f_0(980)$ peak. See the text for details.}
\end{figure}

Such a fit gives, in remarkable agreement with the prediction of
Ref. \cite{ADS1}, the negligible values of the direct coupling
constants $g^{(0)}_{\sigma\gamma\gamma}$ and
$g^{(0)}_{f_0\gamma\gamma}$:\, $\Gamma^{(0)}_{\sigma\to\gamma
\gamma}(m^2_{\sigma})=|m^2_\sigma
g^{(0)}_{\sigma\gamma\gamma}|^2/(16\pi m_\sigma)=0.005$ keV and
$\Gamma^{(0)}_{f_0\to\gamma \gamma}(m^2_{f_0})=|m^2_{f_0}
g^{(0)}_{f_0\gamma\gamma}|^2/(16\pi m_{f_0})=0.00007$ keV; here
$m_{f_0}=0.972$ GeV \cite{FN3}. Note, for comparison, that according
to estimates presented in Refs. \cite{AS1,AS2} the
$\sigma(600)\to\gamma \gamma$ decay width via the $\pi^+\pi^-$ loop
mechanism is of about $1-2$ keV for $0.4<\sqrt{s}<0.5$ GeV
\cite{AS2}, and the $f_0(980)\to\gamma\gamma$ decay width via the
$K^+K^-$ loop mechanism, averaged by the $f_0(980)$ resonance mass
distribution in the $\pi\pi$ channel, is of about $0.15-0.2$ keV
\cite{AS1}. The results of fitting the Belle data are shown in Figs.
2(a) and 2(b). For comparison, Figs. 2(c) and 2(d) demonstrate the
curves corresponding to the fit in which the Crystal Ball data are
not taken into account and $g^{(0)}_{\sigma\gamma\gamma}$\,=\,0. In
this case, $\Gamma^{(0)}_{f_0\to\gamma \gamma}(m^2_{f_0})=0.002$ keV
and $m_{f_0}=0.97$ GeV. The fits to the Belle data for
$g^{(0)}_{\sigma\gamma\gamma}$\,=\,$g^{(0)}_{f_0\gamma\gamma}$\,=\,0
are shown in Figs. 2(e) and 2(f); here $m_{f_0}=0.97$ GeV.
Corresponding curves describing the Crystal Ball data in the region
$2m_\pi<\sqrt{s}<0.8$ GeV are not shown because, for the above three
variants, they practically coincide with each other and with the
curve in Fig. 1(d).

\begin{figure} \includegraphics{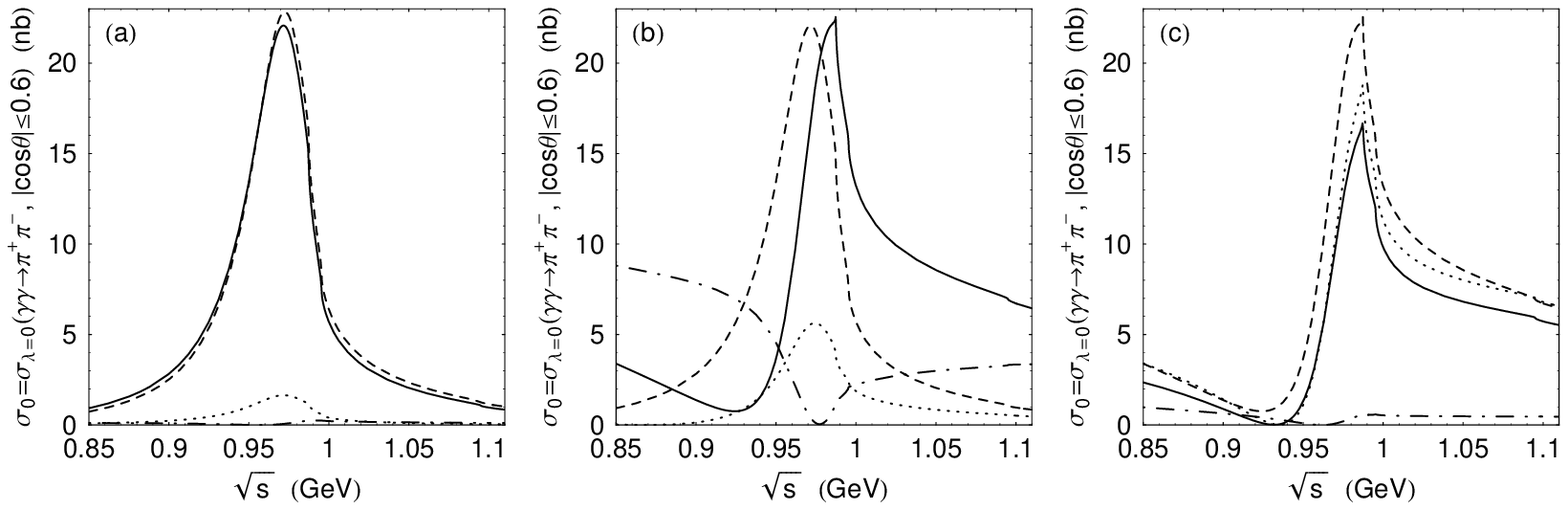} \caption {The main components
shaping the $f_0(980)$ signal in $\sigma_0$. The contributions from
the $\gamma\gamma\to K^+K^-\to\pi^+\pi^-$ and
$\gamma\gamma\to\pi^+\pi^-\to\pi^+\pi^-$ transition amplitudes and
from the Born and direct $\gamma\gamma$ resonance decay amplitudes
are presented. The detailed description of the plotted curves is
given in the text.}\end{figure}

Thus, the available data on $\gamma\gamma\to\pi\pi$ tell us that the
direct couplings of the $\sigma(600)$ and $f_0(980)$ to
$\gamma\gamma$ seem to be very small and that the
$\sigma(600)\to\gamma\gamma$ and $f_0(980)\to\gamma\gamma$ decays
are in fact the four-quark transitions, because they are totally
dominated by the $\pi^+\pi^-$ and $K^+K^-$ loop mechanisms,
respectively.

To gain a complete understanding of the $f_0(980)$ production
mechanism in our model, we present in Figs. 3(a)--3(c) all the main
components shaping the $f_0(980)$ signal in $\sigma_0$ [see Eqs.
(\ref{CS0pm}) and (\ref{CS0})] using the fit shown in Figs. 2(a) and
2(b). Above all, we note that all the curves plotted in Figs. 3(a)
and 3(b) correspond to the different contributions to $\sigma_0$
from the first term in the curly brackets in Eq. (\ref{CS0pm}),
$Z_0|\widetilde{A}_S(s)|^2$ with $Z_0=0.6$, that is, only from the
contributions caused by the final-state interactions. The resulting
picture involving the Born and direct $\gamma\gamma$ resonance decay
contributions [see Eq. (\ref{MSpm}), (\ref{CS0pm}), (\ref{CS0}), and
\ref{Mrd})] is depicted in Fig. 3(c).

The crucial contribution from the $\gamma\gamma\to
K^+K^-\to\pi^+\pi^-$ transition amplitude,
$\widetilde{A}_S(s)=8\alpha I_{K^+K^-}(s)T_{K^+K^-\to\pi^+\pi^-}(s)$
[see Eq. (\ref{MSpm}), (\ref{TKpi}), and (\ref{TKpir})], to
$\sigma_0$ is shown by the solid curve in Fig. 3(a). This
contribution provides the natural scale of the resonance structure
in $\sigma_0$ in the 1 GeV region. The other curves in the figure
represent its constituents. The dashed (dot-dashed) curve
corresponds to the contribution from the last (first) two terms in
the numerator of Eq. (\ref{TKpir}) for
$T^{K^+K^-\to\pi^+\pi^-}_{\mbox{\scriptsize{res}}}(s)$, i.e., only
from the $f_0$ ($\sigma$) production in the $K^+K^-$ channel; see
Ref. \cite{FN4}. The dotted curve corresponds to the contributions
from the last term in the numerator of Eq. (\ref{TKpir}), i.e., from
the $K^+K^-\to f_0\to\sigma\to\pi^+\pi^-$ transition amplitude
caused by the $f_0-\sigma$ mixing.

The dotted curve in Fig. 3(b) shows the contribution to $\sigma_0$
from $\widetilde{A}_S(s)=8\alpha
I_{\pi^+\pi^-}(s)e^{2i\delta^{\pi\pi}_{B}(s)}
2T^{\pi\pi}_{\mbox{\scriptsize{res}}}(s)/3$, i.e, from the $S$ wave
resonance part of the $\gamma\gamma\to\pi^+\pi^-\to\pi^+\pi^-$
transition amplitude [see Eqs. (\ref{MSpm}), (\ref{CS0pm}),
(\ref{CS0}), (\ref{Tpipi}), (\ref{T00}),  and (\ref{Tpipir})]. The
dot-dashed curve in Fig. 3(b) corresponds to the contribution from
$\widetilde{A}_S(s)=8\alpha I_{\pi^+\pi^-}(s)T_{\pi^+\pi^-\to
\pi^+\pi^-}(s)$, i.e, from the full $S$ wave
$\gamma\gamma\to\pi^+\pi^-\to\pi^+\pi^-$ transition amplitude [see
also Eqs. (\ref{MSpm}), (\ref{CS0pm}), (\ref{CS0}), (\ref{Tpipi}),
(\ref{T00}),  and (\ref{Tpipir})]. As for the dashed curve in Fig.
3(b), it is identical to the solid one in Fig. 3(a) and is shown for
direct comparison of the $\gamma\gamma\to K^+K^-\to\pi^+\pi^-$ and
$\gamma\gamma\to\pi^+\pi^-\to\pi^+\pi^-$ contributions. Finally, the
solid curve in Fig. 3(b) shows the total contribution to $\sigma_0$
from the $\gamma\gamma\to K^+K^-\to\pi^+\pi^-$ and
$\gamma\gamma\to\pi^+\pi^-\to\pi^+\pi^-$ rescattering amplitudes,
$\widetilde{A}_S(s)=8\alpha
I_{K^+K^-}(s)T_{K^+K^-\to\pi^+\pi^-}(s)$\,+\,$8\alpha
I_{\pi^+\pi^-}(s)T_{\pi^+\pi^-\to \pi^+\pi^-}(s)$. A comparison of
the dashed, dot-dashed, and solid curves in the figure gives a good
idea of the important role of the interference between the
background and resonance contributions.

The dashed curve in Fig. 3(c) shows the total contribution to
$\sigma_0$ from the $\gamma\gamma\to K^+K^-\to\pi^+\pi^-$ and
$\gamma\gamma\to\pi^+\pi^-\to\pi^+\pi^-$ rescattering amplitudes;
i.e., it is identical to the solid curve in Fig. 3(b). The dotted
curve in Fig. 3(c) corresponds to the contribution from the above
rescattering amplitudes plus the Born contributions [see Eqs.
(\ref{MSpm}), (\ref{CS0pm}), and (\ref{CS0})], and the solid curve
in the figure represents the resulting  picture of the $f_0(980)$
resonance manifestation in $\gamma\gamma\to\pi^+\pi^-$, taking into
account the contribution from the direct
$\sigma(600)\to\gamma\gamma$ and $f_0(980)\to\gamma\gamma$ decays,
see Eq. (\ref{Mrd}) and also Ref. \cite{FN5}.

We finish this section with a general comment. As already emphasized
in Ref. \cite{AS2}, the complex residues of the $\sigma$ pole in the
$\pi\pi\to\pi\pi$ and $\gamma\gamma\to \pi\pi$ amplitudes, used to
estimate the $\sigma\to\gamma\gamma$ decay width \cite{Pe}, do not
give us an idea about the nature of the $\sigma$ meson. Furthermore,
as noted in Ref. \cite{A1}, the majority of current investigations
of the mass spectra in scalar channels do not study particle
production mechanisms. Because of this, such investigations are
essentially preprocessing experiments, and  the derivable
information is very relative. For example, the very first estimate
of the $\sigma$ coupling to the photons via a two-pion intermediate
state \cite{DG} was restricted to the case of the ``bare'' (without
any background) $\sigma$ meson, which contradicts the low energy
chiral dynamics, and the recent estimates of the
$f_0(980)\to\gamma\gamma$ decay width \cite{B2,B3} have not taken
into account the rapidly changing $K^+K^-$ loop production mechanism
of the $f_0(980)$ \cite{AS1}. Nevertheless, the progress in
understanding the particle production mechanisms could essentially
help us reveal the light scalar meson nature.

\section{THE \lowercase{{\boldmath $f_2(1270)$}} RESONANCE CONTRIBUTION}

To estimate the $f_2(1270)\to\gamma\gamma$ decay width,
$\Gamma_{f_2\to\gamma\gamma}(m^2_{f_2})$, from the data on
$\gamma\gamma\to\pi\pi$, it is usually assumed that the $f_2(1270)$
decay occurs mostly into $\gamma\gamma$ states with $\lambda=2$
\cite{E1,E2,E4,B3,Jo}. For this, the specific models for the
background amplitudes with $\lambda$\,=\,2 and 0 are also needed
\cite{E1,E2,E4,B3, Jo}. The large background under the $f_2(1270)$
in the $\gamma\gamma\to\pi^+\pi^-$ channel [see, for example, Fig.
1(c)] is dominated by the Born amplitude with $\lambda$\,=\,2. The
background situation in the $\gamma\gamma\to\pi^0\pi^0$ channel is
more pure. Here, however, uncertainties in the data and bins of
$\sqrt{s}$ are still rather large. Different assumptions about the
background amplitudes in the $f_2(1270)$ region have been used in
the literature \cite{E1,E2,E4,B3,Jo}. In so doing, the central
values of $\Gamma_{f_2\to\gamma\gamma}(m^2_{f_2})$ obtained in the
independent experiments lie in the range from 2.3 to 3.6 keV
\cite{E1,E2,E4,Jo,PDG}.

According to the Particle Data Group (PDG) estimate \cite{PDG},
$\Gamma_{f_2\to\gamma\gamma}(m^2_{f_2})=2.6 \pm0.24$ keV. In the
recent work \cite{B3}, the authors of the Belle experiment presented
``a consistency check'' of their data in the $f_2(1270)$ region with
this estimate. They fixed the values of the $f_2(1270)$ resonance
parameters as given by the PDG \cite{PDG} and, using a simple
phenomenological parametrization of the background amplitudes,
performed the fit to the data on $\sigma(\gamma\gamma\to\pi^+\pi^-;
|\cos\theta|\leq0.6)$ in the region $m_{f_2}-\Gamma^{tot}_{f_2}\leq
\sqrt{s}\leq m_{f_2}+\Gamma^{tot}_{f_2}$, i.e., for
$1.090\leq\sqrt{s}\leq1.461$ GeV. The resulting fit turned out to be
very good, and they concluded that the ``consistency check is
satisfactory.'' Unfortunately, in Ref. \cite{B3} the factor
$\sqrt{2/3}$ has been missed in the $f_2(1270)$ production amplitude
in Eq. (10). Thus, the consistency is broken. In fact, it follows
from the Belle data \cite{B3} that $\Gamma_{f_2\to\gamma\gamma}
(m^2_{f_2})\approx3.9$ keV, which is 1.5 times greater than the PDG
estimate. At the same time, this value is in close agreement with
our estimate, $\Gamma_{f_2\to\gamma\gamma} (m^2_{f_2})\approx3.8$
keV, which we obtained from the Belle data, but in another way (see
below).

To describe the Belle data in the region $0.8\leq\sqrt{s}\leq1.5$
GeV, we use the expression for the total cross section
$\sigma(\gamma\gamma\to\pi^+\pi^-;|\cos\theta|\leq0.6)=\sigma_0+\sigma_2$.
The cross section $\sigma_0$ has been constructed in Secs. III and
IV,  and the cross section $\sigma_2$ involving the Born and
$f_2(1270)$ resonance contributions has the form \cite{Ly,Jo}:
\begin{equation}\label{CS2pm}
\sigma_2=\frac{8\pi}{s}\int^{0.6}_0
\left|\,\frac{\sqrt{\rho_{\pi^+}(s)}}{16\pi}\,M_2^{\mbox{\scriptsize{Born}}}
(s,\theta)\,+\,5\,d^2_{20}(\theta)\,\frac{\sqrt{s}\,G_2(s) \sqrt{2
\Gamma_{f_2\to\pi\pi}(s)/3}}{m^2_{f_2}-s-i\sqrt{s}\Gamma^{\mbox{\scriptsize{tot}}
}_{f_2}(s)}\,\right|^2d\cos\theta\,,
\end{equation} where
$M_2^{\mbox{\scriptsize{Born}}}(s,\theta)=8\pi\alpha\,\rho^2_{\pi^+}(s)\,
\sin^2\theta/[1-\rho^2_{\pi^+}(s)\,\cos^2\theta]$ is the Born
helicity-2 amplitude $\gamma\gamma\to\pi^+\pi^-$,\,
$d^2_{20}(\theta)=\frac{\sqrt{6}}{4}\sin^2\theta$, and the
energy-dependent total width of the $f_2(1270)$ is given by
$\Gamma^{\mbox{\scriptsize{tot}} }_{f_2}(s)
$\,=\,$\Gamma_{f_2\to\pi\pi}(s)+\Gamma_{f_2\to K\bar
K}(s)+\Gamma_{f_2\to4\pi}(s)$. The partial width
$\Gamma_{f_2\to\pi\pi}(s)$ is parameterized as \cite{E1}
\begin{equation}\label{Wf2pipi}
\Gamma_{f_2\to\pi\pi}(s)=\Gamma^{\mbox{\scriptsize{tot}}
}_{f_2}(m^2_{f_2})B(f_2\to\pi\pi)\frac{m^2_{f_2}}{s}\frac{q^5_{\pi^+}(s)}{q^5_{\pi^+}(m^2_{f_2})}
\frac{D_2(q_{\pi^+}(s)R_{f_2})}{D_2(q_{\pi^+}(m^2_{f_2})R_{f_2})}\,,\end{equation}
where $D_2(x)$\,=\,$1/(9+3x^2+x^4)$,
$q_{\pi^+}(s)$\,=\,$\sqrt{s}\rho_{\pi^+}(s)/2$, and $R_{f_2}$ is an
interaction radius. $\Gamma_{f_2\to K\bar K}(s)$ has the form
similar to Eq. (17). $\Gamma_{f_2\to4\pi}(s)$ as a function of $s$
is approximated by the $S$ wave $f_2(1270)\to\rho\rho\to4\pi$ decay
width; see, for example, Ref. \cite{ADS1}. The branching ratios are
$B(f_2\to\pi\pi)$\,=\,0.847, $B(f_2\to K\bar K)$\,=\,0.046, and
$B(f_2\to4\pi)$\,=\,0.107 \cite{PDG}. Finally,
\begin{equation}\label{G2}
G_2(s)=\sqrt{\Gamma^{(0)}_{f_2\to\gamma\gamma}(s)}+i\frac{\sqrt{\rho_{\pi^+}(s)}}{16\pi}
\,M_{22}^{\mbox{\scriptsize{Born}}}(s)\sqrt{
2\Gamma_{f_2\to\pi\pi}(s)/3}\,.
\end{equation} By definition, $\Gamma_{f_2\to\gamma\gamma} (s)=|G_2(s)|^2$.
For $\Gamma^{(0)}_{f_2\to\gamma\gamma}(s)$ we use the following
parametrization:
\begin{equation}\label{Wf20}
\Gamma^{(0)}_{f_2\to\gamma\gamma}(s)=\frac{m_{f_2}}{\sqrt{s}}
\Gamma^{(0)}_{f_2\to\gamma\gamma}(m^2_{f_2})
\left(\frac{s^2}{m^4_{f_2}}\,\frac{m^2_{f_2}+\Lambda^2_{f_2}}{s+\Lambda^2_{f_2}}\right)^2\,.
\end{equation}
The second term in Eq. (\ref{G2}) corresponds to the four-quark
transition of the $f_2(1270)$ into photons via the $\pi^+\pi^-$ real
intermediate state, $f_2(1270)\to\pi^+\pi^-\to\gamma\gamma$, where
\begin{equation}\label{M22B}
M_{22}^{\mbox{\scriptsize{Born}}}(s)=4\pi\alpha\sqrt{\frac{3}{2}}\left[\frac{1-\rho^2_{\pi^+}(s)}
{2\rho^3_{\pi^+}(s)}\ln\frac{1+\rho_{\pi^+}(s)}{1-\rho_{\pi^+}(s)}-\frac{1}{\rho^2_{\pi^+}(s)}+
\frac{5}{3}\right]
\end{equation}
is the Born partial, helicity amplitude $\gamma\gamma\to\pi^+\pi^-$
with $J=\lambda=2$. This term ensures the fulfilment of the Watson
theorem requirement for the $I=0$, $J=\lambda=2$ amplitude
$\gamma\gamma\to\pi\pi$ in the elastic region. It gives a rather
small contribution to $\Gamma_{f_2\to\gamma\gamma}(m^2_{f_2})$:
\begin{equation}\label{Wf2gg}\Gamma_{f_2\to\gamma\gamma}(m^2_{f_2})=
\Gamma^{(0)}_{f_2\to\gamma\gamma}(m^2_{f_2})+0.21\,\mbox{keV}\,.
\end{equation}
It is generally accepted that the $f_2(1270)\to\gamma\gamma$ decay
rate is dominated by the direct quark-antiquark transition $q\bar
q\to\gamma\gamma$, that is, by
$\Gamma^{(0)}_{f_2\to\gamma\gamma}(m^2_{f_2})$. As we saw above, for
the lightest scalar mesons, the situation is reversed.

In the fit, we use as free parameters
$\Gamma^{(0)}_{f_2\to\gamma\gamma}(m_{f_2})$, $m_{f_2}$,
$\Gamma^{\mbox{\scriptsize{tot}}}_{f_2}(m^2_{f_2})$, $R_{f_2}$,
$\Lambda_{f_2}$, and also $g^{(0)}_{\sigma\gamma\gamma}$,
$g^{(0)}_{f_0\gamma\gamma}$, and $m_{f_0}$. The parameters $R_{f_2}$
and $\Lambda_{f_2}$ control the $s$ dependencies of the $f_2(1270)$
partial decay widths, and consequently, they are responsible for the
shape of the $f_2(1270)$ line. The results of the fit to the Belle
data are shown in Figs. 4(a)--4(c). Figure 4(a) demonstrates the
overall picture of the most important contributions in the region
$2m_\pi\leq\sqrt{s}\leq1.5$ GeV; here $\sigma_{f_2}$ is the cross
section corresponding to the $f_2(1270)$ resonance contribution and
the dotted curve shows the contribution of the interference between
the $f_2(1270)$ and background amplitudes in $\sigma_2$. The
descriptions of the Belle data in the whole investigated region of
$\sqrt{s}$ and in the $f_0(980)$ resonance region are presented in
more detail in Figs. 4(b) and 4(c), respectively. The parameters
obtained are $\Gamma^{(0)}_{f_2\to\gamma\gamma}(m_{f_2})=3.59$ keV
[$\Gamma_{f_2\to\gamma\gamma}(m_{f_2})=3.8$ keV, see Eq.
(\ref{Wf2gg})], $m_{f_2}$\,=\,1.279\,GeV,
$\Gamma^{\mbox{\scriptsize{tot}}}_{f_2}(m^2_{f_2})=0.188$ GeV,
$R_{f_2}=2.575$ GeV$^{-1}$, $\Lambda_{f_2}$\,=\,3.52\,GeV,
$g^{(0)}_{\sigma\gamma\gamma}$\,=\,0.482\,GeV$^{-1}$
[$\Gamma^{(0)}_{\sigma\to\gamma
\gamma}(m^2_{\sigma})$\,=\,0.01\,keV],
$g^{(0)}_{f_0\gamma\gamma}$\,=\,-0.867\,GeV$^{-1}$
[$\Gamma^{(0)}_{f_0\to\gamma \gamma}(m^2_{f_0})$\,=\,0.015\,keV],
and $m_{f_0}$\,=\,0.975\,GeV. Because of the smallness of
statistical uncertainties in the Belle data, the formally calculated
errors of the above-listed parameters turn out to be negligible. In
similar situations, the model dependence of the fitted parameter
values is the most important source of their uncertainty.

\begin{figure} \includegraphics{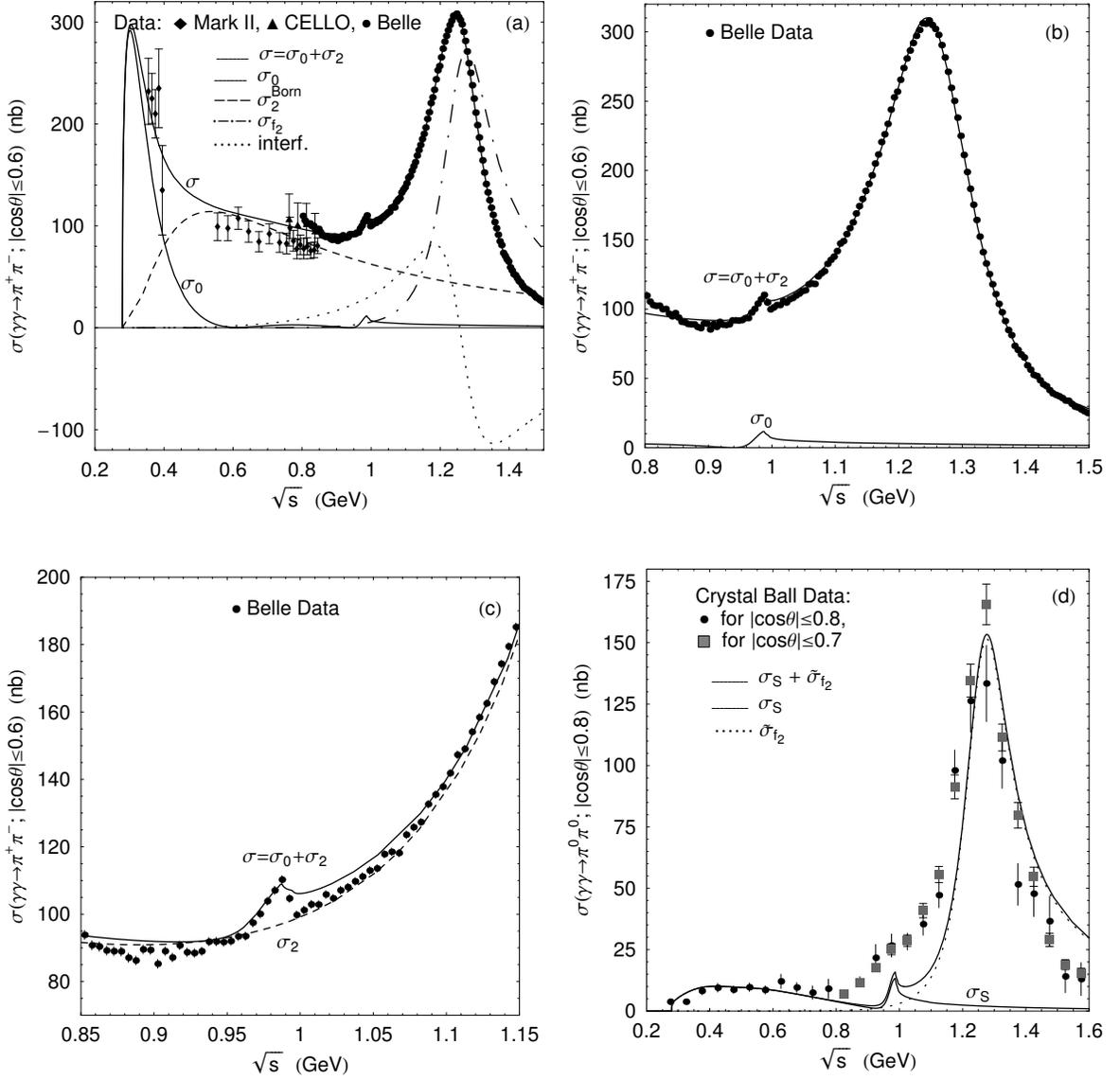} \caption {The fit to the Belle
data on the  $\gamma\gamma\to\pi^+\pi^-$ cross section and the
comparison with the Crystal Ball data on the
$\gamma\gamma\to\pi^0\pi^0$ cross section. See the text for
details.}\end{figure}

The obtained description of the Belle data as a whole seems to be
quite satisfactory, except for minor details \cite{FN6}. The more
important result of the fit is that the values of the direct
coupling constants of the $\sigma(600)$ and $f_0(980)$ resonances to
$\gamma\gamma$ turn out to be small (see the above-mentioned
corresponding decay widths). Of course, the obtained concrete values
of these constants are, evidently, rather conditional (compare, for
example, the above fitting variant with that presented in Sec. IV).
Let us stress, however, that the very fact of the suppression of the
direct $\sigma(600)$ and $f_0(980)$ couplings to photons
(corresponding widths are much less than 1 keV) can be considered to
be well established. In addition, it appears from the new Belle data
that $\Gamma_{f_2\to\gamma\gamma}(m_{f_2})$ is about a factor of 1.5
higher than the estimate quoted by the PDG \cite{PDG}.

We now construct the $\gamma\gamma\to\pi^0\pi^0$ reaction cross
section $\sigma(\gamma\gamma\to\pi^0\pi^0;|\cos\theta|\leq0.8)$\,
=\,$\sigma_S+\tilde{\sigma}_{f_2}$, where $\tilde{\sigma}_{f_2}$ is
the $f_2(1270)$ production cross section in the $\pi^0\pi^0$ channel
(an analog of $\sigma_{f_2}$ for the $\pi^+\pi^-$ one), and compare
it with the Crystal Ball data \cite{E1,E5}. The result is shown in
Fig. 4(d). As is seen, the agreement with the data is very poor in
the whole region of the $f_2(1270)$ resonance influence, i.e., for
$\sqrt{s}$ from 0.8 to 1.6 GeV. We verified that the parametrization
of the $f_2(1270)$ contribution used by the Belle Collaboration
\cite{B3} leads to a similar resonance pattern. To improve the
description of the available data on the reaction
$\gamma\gamma\to\pi^0\pi^0$, it is necessary to raise the left wing
of the $f_2(1270)$ resonance and to lower its right one \cite{FN7},
i.e., to change the $f_2(1270)$ resonance shape in comparison with
that established from the $\gamma\gamma\to\pi^+\pi^-$ data.

As for the above difficulty, in fact, it arose as the first detailed
experiments were carried out on the reactions
$\gamma\gamma\to\pi^0\pi^0$ \cite{E1,E5} and
$\gamma\gamma\to\pi^+\pi^-$ \cite{E2,E4}. Different parametrizations
of the $f_2(1270)$ resonance shape have been used in Refs.
\cite{E1,E2,E4}. Taking the corresponding formulas and fitting
parameters from Refs. \cite{E1,E2,E4}, we have made sure that in the
early analyses \cite{E1,E2,E4} the appreciably different shapes of
the $f_2(1270)$ peak in the $\pi^0\pi^0$ and $\pi^+\pi^-$ channels
were obtained. The difference bears the above-mentioned character.
However, the existing uncertainties in the Crystal Ball
\cite{E1,E5}, Mark II \cite{E2}, and CELLO \cite{E4} data [see, for
example, Figs. 1(d) and 1(a)] hamper any definite conclusions about
their possible inconsistency or about the urgent need for searching
the additional mechanisms to obtain a good simultaneous description
of the $\pi^0\pi^0$ and $\pi^+\pi^-$ data in the $f_2(1270)$ region.
It is clear that the Belle experiment on the reaction
$\gamma\gamma\to\pi^+\pi^-$ \cite{B2,B3} essentially aggravates the
situation because its results are based on statistics which are
about 3 orders of magnitude higher than those collected in the
Crystal Ball experiments. So, it is clear that, in the first place,
very high quality data on the reaction $\gamma\gamma\to\pi^0\pi^0$
and their partial wave analysis would be extremely useful to obtain
reliable conclusions from the simultaneous description of the
$\pi^+\pi^-$ and $\pi^0\pi^0 $ channels \cite{FN8}.

After this work was completed, very high statistics Belle data on
the reaction $\gamma\gamma\to\pi^0\pi^0$ for $\sqrt{s}>0.6$ GeV
\cite{Abe} appeared, which are in close agreement with the Crystal
Ball measurements \cite{E1,E5}. Probably, this implies that a
damping form factor \cite{FN8} in the $\gamma\gamma\to\pi^+\pi^-$
Born amplitudes is really needed for the simultaneous description of
the $\pi^+\pi^-$ and $\pi^0\pi^0$ production cross sections in the
$f_2(1270)$ resonance region. Such an investigation requires
considerable efforts, because the form factor influence should be
taken into account in partial waves and in loop contributions. We
shall present a careful analysis of the compatibility of the new
$\gamma\gamma\to\pi^+\pi^-$ and $\gamma\gamma\to\pi^0\pi^0$ data
from Belle elsewhere, together with comments on the choice of a
suitable phenomenological form factor. Here we only announce some
preliminary results of our analysis.

New high statistics results from Belle on the
$\gamma\gamma\to\pi^0\pi^0$ reaction cross section \cite{Abe} are
shown in Fig. 5. In spite of very small statistical errors
\cite{Abe}, these data can be quite satisfactorily described
\cite{FN7}, separately from the data for the
$\gamma\gamma\to\pi^+\pi^-$ production; see, as an example, the
solid curve in Fig. 5(b) which corresponds to the parameters
$\Gamma^{(0)}_{f_2\to\gamma\gamma}(m_{f_2})=3.24$ keV
[$\Gamma_{f_2\to\gamma\gamma}(m_{f_2})=3.43$ keV],
$m_{f_2}$\,=\,1.272\,GeV,
$\Gamma^{\mbox{\scriptsize{tot}}}_{f_2}(m^2_{f_2})=0.183$ GeV,
$R_{f_2}=6.5$ GeV$^{-1}$, $\Lambda_{f_2}$\,=\,0,
$g^{(0)}_{\sigma\gamma\gamma}$\,=\,0.542\,GeV$^{-1}$,
$g^{(0)}_{f_0\gamma\gamma}$\,=\,0.468\,GeV$^{-1}$, and
$m_{f_0}$\,=\,0.969\,GeV. However, such a fit to the
$\gamma\gamma\to\pi^0\pi^0$ cross section is in apparent
contradiction with the $\gamma\gamma\to\pi^+\pi^-$ data, see the
solid curve for $\sigma=\sigma_0+\sigma_2$ in Fig. 5(a). This is
caused mainly by the large Born contributions to $\sigma_2$ in the
$\gamma\gamma\to\pi^+\pi^-$ channel. Recall that such contributions
are absent in $\gamma\gamma\to\pi^0\pi^0$. Thus, the situation can
be essentially improved by multiplying the
$\gamma\gamma\to\pi^+\pi^-$ Born amplitudes by some overall, damping
form factor $G(t,u)$ \cite{FN8,Po}, where $t$ and $u$ are the usual
Mandelstam variables for the reaction $\gamma\gamma\to\pi^+\pi^-$.
For this we use here the expression proposed by Poppe \cite{Po},
$$G(t,u)=-\frac{1}{s}\left[\frac{t-m^2_{\pi^+}}{1-(u-m^2_{\pi^+})/x^2_1}+
\frac{u-m^2_{\pi^+}}{1-(t-m^2_{\pi^+})/x^2_1}\right]\,,$$ where
$x_1$ is a free parameter. This ansatz is quite acceptable in the
physical region of the reaction $\gamma\gamma\to\pi^+\pi^-$.
Replacing $m_{\pi^+}$ and $x_1$ by $m_{K^+}$ and $x_2$,
respectively, we also obtain a form factor suitable for the
$\gamma\gamma\to K^+K^-$ Born amplitude. The solid curves for
$\sigma=\sigma_0+\sigma_2$ and $\sigma_S+\tilde{\sigma}_{f_2}$ in
Figs. 5(c) and 5(d), respectively, show an example of the overall
fit to the new $\gamma\gamma\to\pi^+\pi^-$ and
$\gamma\gamma\to\pi^0\pi^0$ data, taking into account the form
factors modifying the Born contributions. The obtained description
is quite reasonable (if not excellent), but only within systematic
errors of the data, which are plotted in Figs. 5(c) and 5(d) in the
form of the shaded bands. We think that such a treatment of the high
statistics Belle data is sufficiently justified. Statistical errors
of the two Belle measurements are so small that obtaining the
formally acceptable $\chi^2$  for simultaneous fits to the data is
practically impossible. The curves in Figs. 5(c) and 5(d) correspond
to the parameters $\Gamma^{(0)}_{f_2\to\gamma\gamma}(m_{f_2})=3.60$
keV [$\Gamma_{f_2\to\gamma\gamma}(m_{f_2})=3.68$ keV; we consider
this estimate for $\Gamma_{f_2\to\gamma\gamma}(m_{f_2})$ as the most
preferable one], $m_{f_2}$\,=\,1.272\,GeV,
$\Gamma^{\mbox{\scriptsize{tot}}}_{f_2}(m^2_{f_2})=0.188$ GeV,
$R_{f_2}=5$ GeV$^{-1}$, $\Lambda_{f_2}$\,=\,0,
$g^{(0)}_{\sigma\gamma\gamma}= g^{(0)}_{f_0\gamma\gamma}=0$,
$m_{f_0}$\,=\,0.969\,GeV, $x_1=1$\, GeV, and $x_2=3$\, GeV. A
comparison of Figs. 5(b) and 5(d) shows that the impact of the form
factor on the $\gamma\gamma\to\pi^0\pi^0$ cross section turns out to
be really small, in contrast to the case of the
$\gamma\gamma\to\pi^+\pi^-$ production; see Figs. 5(a) and 5(c).
Note also that all the above conclusions about the production
mechanisms of the $\sigma(600)$ and $f_0(980)$ resonances and a
comment on the angular distribution for the nonresonance background,
given in Sec. II, remain valid.

\begin{figure} \includegraphics{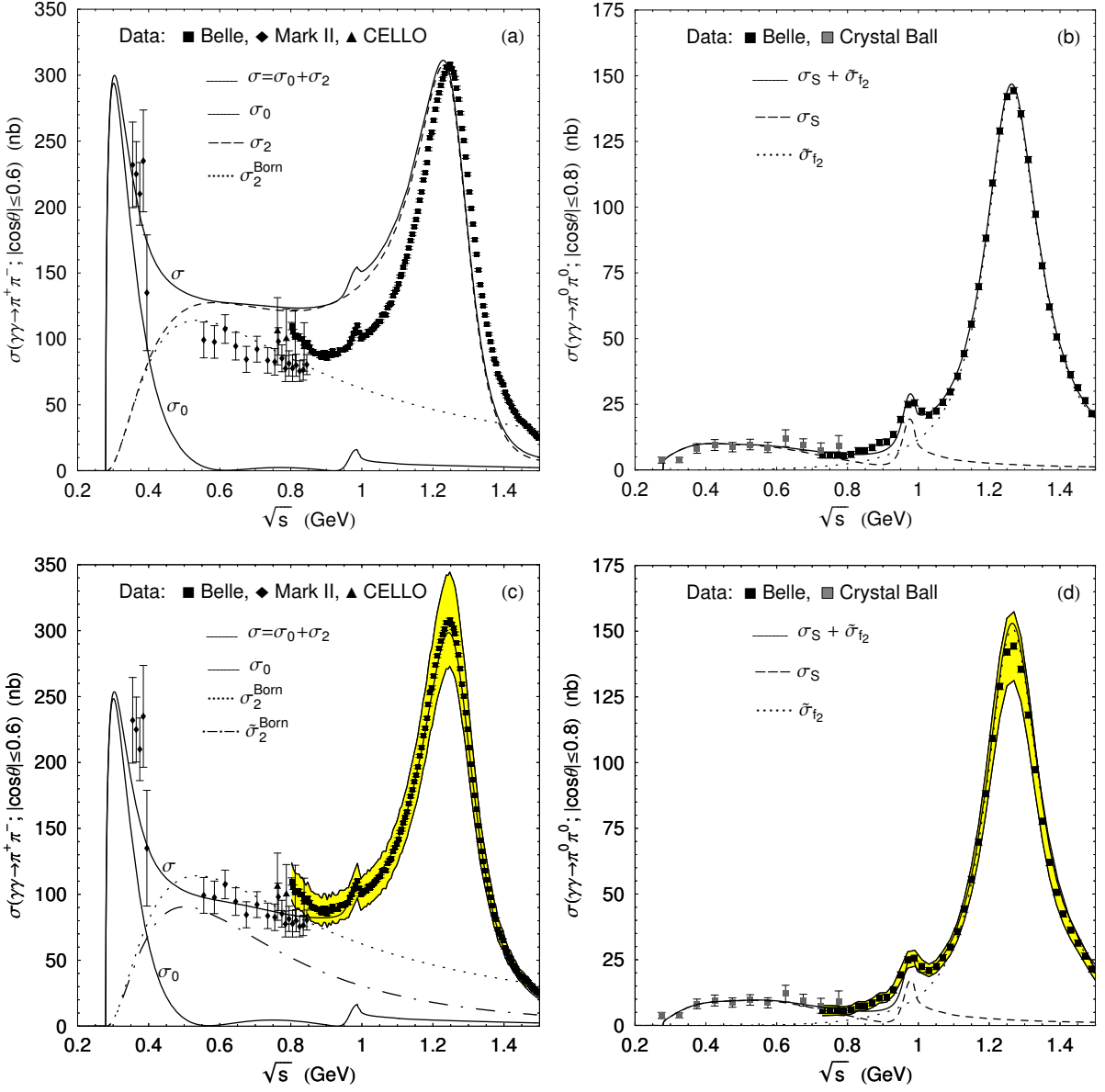} \caption {An illustration of the
simultaneous description of the new high statistics Belle data on
the $\gamma\gamma\to\pi^+\pi^-$ \cite{B2,B3} and
$\gamma\gamma\to\pi^0\pi^0$ \cite{Abe} reaction cross sections. The
Belle data in plots (a), (b), (c), and (d) are represented by full
squares with statistical error bars. The shaded bands in (c) and (d)
correspond to the Belle data taking into account their systematic
errors \cite{B2,B3,Abe}. The curves in (b) and (a) correspond to the
fit to the $\gamma\gamma\to\pi^0\pi^0$ data and its consequence for
the $\gamma\gamma\to\pi^+\pi^-$ channel, respectively. The curves in
(c) and (d) correspond to the overall fit to the
$\gamma\gamma\to\pi^+\pi^-$ and $\gamma\gamma\to\pi^0\pi^0$ data in
the regions $0.85<\sqrt{s}<1.5$ GeV and  $2m_\pi<\sqrt{s}<1.5$ GeV,
respectively, in the model with a form factor.
$\tilde{\sigma}^{\mbox{\scriptsize{Born}}}_2$ in (c) is the
$\lambda=2$ Born cross section modified by a form factor. See the
text for details.}\end{figure}

\section{CONCLUSION}

We have analyzed the new high statistics Belle data on the reaction
$\gamma\gamma\to\pi^+\pi^-$ taking into account its main dynamical
mechanisms. The analysis has shown that the direct coupling
constants of the $\sigma(600)$ and $f_0(980)$ resonances to
$\gamma\gamma$ are small, which is typical \cite{ADS1} for the
four-quark nature \cite{Ja} of these states. Our main conclusion is
that the $\sigma(600)\to\gamma\gamma$ and $f_0(980)\to\gamma\gamma$
decays are the four-quark transitions that are dominated mainly by
the $\pi^+\pi^-$ and $K^+K^-$ loop mechanisms, respectively. In
addition, we have presented some results of a simultaneous
description of the $\gamma\gamma\to\pi^+\pi^-$ data and the latest,
very high statistics Belle data on the reaction
$\gamma\gamma\to\pi^0\pi^0$. We have also estimated the
$f_2(1270)\to\gamma\gamma$ decay width. We intend to develop the
above analysis to further understand the light scalar meson physics.

\begin{center}{\bf ACKNOWLEDGMENTS} \end{center}

We thank Y. Watanabe very much for communication and for the table
with the Belle data for $\gamma\gamma\to\pi^+\pi^-$. This work was
supported in part by the Presidential Grant No. NSh-5362.2006.2 for
Leading Scientific Schools and by the RFFI Grant No. 07-02-00093
from Russian Foundation for Basic Research.


\begin{thebibliography}{99}
\bibitem{B1}  T. Mori {\it et al.} (Belle Collaboration), in
              {\it Proceedings of the International
              Simposium on Hadron Spectroscopy Chiral Symmetry and
              Relativistic Description of Bound Systems, Tokyo, 2003},
              edited by S. Ishida, K. Takamatsu, T. Tsuru, S.Y. Tsai,
              M. Ishida, and T. Komada (KEK Proceedings 2003-7) (KEK,
              Tsukuba, 2003), p. 159.
\bibitem{B2}  T. Mori {\it et al.} (Belle Collaboration), Phys. Rev. D {\bf 75}, 051101 (2007).
\bibitem{E1}  H. Marsiske {\it et al.,} Phys. Rev. D {\bf 41}, 3324 (1990).
\bibitem{E2}  J. Boyer {\it et al.,} Phys. Rev. D {\bf 42}, 1350 (1990).
\bibitem{E3}  T. Oest {\it et al.,} Z. Phys. C {\bf 47}, 343 (1990).
\bibitem{E4}  H.J. Behrend {\it et al.,} Z. Phys. C {\bf 56}, 381 (1992).
\bibitem{E5}  J.K. Bienlein, in {\it Proceedings of the IX International
              Workshop on Photon-Photon Collisions, San Diego, 1992},
              edited by D. Caldwell and H.P. Paar (World Scientific,
              1992), p. 241.
\bibitem{E6}  R. Barate {\it et al.,} Phys. Lett. B {\bf 472}, 189 (2000).
\bibitem{ADS1}N.N. Achasov, S.A. Devyanin, and G.N. Shestakov,
              Phys. Lett. {\bf 108B}, 134 (1982); Z. Phys. C {\bf 16}, 55 (1982).
\bibitem{AS1} N.N. Achasov and G.N. Shestakov, Phys. Rev. D {\bf 72}, 013006 (2005);
              Yad. Fiz. {\bf 69}, 1545 (2006) [Phys. At. Nucl. {\bf 69}, 1510 (2006)].
\bibitem{AKS} N.N. Achasov, A.V. Kiselev, and G.N. Shestakov, Nucl. Phys. B Proc. Suppl.
              {\bf 162}, 127 (2006).
\bibitem{A1}  N.N. Achasov, in {\it Proceedings of the 14th International Seminar QUARKS-2006,
              Repino, St. Peterburg, 2006}, edited by S.V. Demidov, V.A. Matveev, V.A. Rubakov,
              and G.I. Rubtsov (INR RAS, Moscow, 2007), p. 37; arXiv: hep-ph/0609261.
\bibitem{AK1} N.N. Achasov and A.V. Kiselev, Phys. Rev. D {\bf 73}, 054029 (2006).
\bibitem{AS2} N.N. Achasov and G.N. Shestakov, Phys. Rev. Lett. {\bf 99}, 072001 (2007);
              arXiv: 0704.2368 [hep-ph].
\bibitem{AS3} N.N. Achasov and G.N. Shestakov, Phys. Rev. D {\bf 49}, 5779 (1994);
              Yad. Fiz. {\bf 56}, No. 9, 206 (1993) [Phys. At. Nucl. {\bf 56}, 1270 (1993)].
\bibitem{B3}  T. Mori {\it et al.} (Belle Collaboration), J. Phys. Soc. Jap. {\bf 76},
              074102 (2007); arXiv: 0704.3538 [hep-ex].
\bibitem{Me}  G. Mennessier, Z. Phys. C {\bf 16}, 241 (1983).
\bibitem{Ly}  D.H. Lyth, J. Phys. G: Nucl. Phys. {\bf 11}, 459 (1985).
\bibitem{Jo}  R.P. Johnson, Ph.D. thesis (SLAC Report No.-294, 1986).
\bibitem{MP}  D. Morgan and M.R. Pennington, Phys. Lett. B {\bf 192}, 207 (1987);
              Z. Phys. C {\bf 37}, 431 (1988); Phys. Lett. B {\bf 272}, 134 (1991).
\bibitem{Pe}  M.R. Pennington, Phys. Rev. Lett. {\bf 97}, 011601 (2006).
\bibitem{AG1} N.N. Achasov and V.V. Gubin, Phys. Rev. D {\bf 57}, 1987 (1998);
              Yad. Fiz. {\bf 61}, 1473 (1998) [Phys. At. Nucl. {\bf 61}, 1367 (1998)].
\bibitem{FN1} The authors of Refs. \cite{B1,B2,B3} erroneously believe that the Flatte work
\cite{Fl} is related to the formula that they  have use in their
works for the $f_0(980)$ propagator $1/D_{f_0}(s)$. This formula for
$1/D_{f_0}(s)$ was first obtained in Ref. \cite{ADS2}.
\bibitem{FN2} There is a misprint in this fit in the sign of the constant
$C\equiv C_{f_0\sigma}$; here we use
$C_{f_0\sigma}=-0.047\,\mbox{GeV}^2$.
\bibitem{AS4} N.N. Achasov and G.N. Shestakov, Phys. Rev. D {\bf 67}, 114018 (2003);
              Yad. Fiz. {\bf 67}, 1380 (2004) [Phys. At. Nucl. {\bf 67}, 1355 (2004)].
\bibitem{MM}  G. Mennessier, P. Minkowski, S. Narison, and W. Ochs, arXiv: 0707.4511 [hep-ph].
\bibitem{FN3} This $m_{f_0}$ value is about 12 MeV smaller than that obtained in fit 1
from Ref. \cite{AK1}, but in so doing the description of the
$\pi\pi$ phase shift $\delta^0_0(s)$ essentially does not
deteriorate.
\bibitem{FN4} This contribution from the $\sigma$ production is negligible, and in Fig. 3(a)
it is increased by a factor of 10.
\bibitem{FN5} The contribution from the direct $\gamma\gamma$ resonance decay amplitude
$M^{\mbox{\scriptsize{direct}}}_{\mbox{\scriptsize{res}}}
(\gamma\gamma\to\pi^+\pi^-;s)$ is very small and that is why this is
increased by a factor of 10 in Fig. 3(c) (see the corresponding
dot-dashed curve). However, as is seen from this figure, the
interference between the amplitude
$M^{\mbox{\scriptsize{direct}}}_{\mbox{\scriptsize{res}}}
(\gamma\gamma\to\pi^+\pi^-;s)$ and the main ones is quite
noticeable.
\bibitem{DG}  P.C. DeCelles and J.F. Goehl, Phys. Rev. {\bf 184}, 1617 (1969).
\bibitem{PDG} W.-M. Yao {\it et al.} (Particle Data Group), J. Phys. G {\bf 33}, 1 (2006).
\bibitem{FN6} In principle, these details [see Figs. 4(b) and
4(c)] can be artifacts associated with the event selection procedure
employed. It must be kept in mind that the systematic errors in the
Belle experiment are an order of magnitude higher than the
statistical ones \cite{B1,B2,B3}. Here we add yet an evident remark.
In order to improve the data description, it is necessary to
increase the flexibility of the phenomenological model in use. As a
rule, this is accompanied both by new assumptions and by introducing
additional free parameters.
\bibitem{FN7} Formally, this is easy
to implement, for example, by increasing $R_{f_2}$ and decreasing
$\Lambda_{f_2}$ and $m_{f_2}$ relative to the $\pi^+\pi^-$ case.
\bibitem{FN8} Generally speaking, the consistent description of
the Crystal Ball and Belle data could be implemented by introducing
a form factor into the Born amplitudes. Such a form factor (see
Refs. \cite{E2,E4,MP,Po,AS5} for details) suppresses the continuum
in the $f_2(1270)$ region in the $\pi^+\pi^-$ channel and, thus,
permits the left wing of the $f_2(1270)$ resonance to be increased
in agreement with the available $\pi^+\pi^-$ and $\pi^0\pi^0 $ data.
\bibitem{Po}  M. Poppe, Int. J. Mod. Phys. A {\bf 1}, 545, (1986).
\bibitem{AS5} N.N. Achasov and G.N. Shestakov, Mod. Phys. Lett. A {\bf 9}, 1351 (1994).
\bibitem{Abe} K. Abe {\it et al.} (Belle Collaboration), arXiv: 0711.1926 [hep-ex].
\bibitem{Ja}  R.L. Jaffe, Phys. Rev. D {\bf 15}, 267, (1977); {\bf 15}, 281 (1977).
\bibitem{Fl}  S.M. Flatte, Phys. Lett. B {\bf 63}, 224 (1976).
\bibitem{ADS2}N.N. Achasov, S.A. Devyanin, and G.N. Shestakov, Yad. Fiz.
              {\bf 32}, 1098 (1980) [Sov. J. Nucl. Phys. {\bf 32}, 566 (1980)].
\end{thebibliography}
\end{document}